\documentstyle[twocolumn,aps,prl]{revtex}
\input epsf
\begin{document}

\title{Remark on formation of colored black holes via fine tuning}

\author{Piotr Bizo\'n${}^{1}$ and Tadeusz Chmaj${}^{2}$}
\address{$^{1}$Insitute of Physics, Jagiellonian University, Krak\'ow,
 Poland}
\address{$^{2}$Institute of Nuclear Physics, Krak\'ow, Poland}

\date{15 June, 1999}

\maketitle

\begin{abstract}
In a recent paper (gr-qc/9903081) Choptuik, Hirschmann, and Marsa have
 discovered
 the scaling
law for the lifetime of an intermediate attractor in the formation 
of n=1 colored black holes via
fine tuning. We show that their result is in agreement with the
prediction of linear perturbation analysis. We also briefly comment on
the dependence of the mass
 gap across the threshold on the radius of the event horizon.

\end{abstract}
\pacs{04.25.Dm, 04.40.-b, 04.70.Bw}
\vskip1pc

\narrowtext
Recently, Choptuik, Hirschmann, and Marsa have discovered a new type
of critical behavior within the black hole regime of the spherically
symmetric Einstein-Yang-Mills model~\cite{chm}. Using the horizon
excision
technique, they were able to follow the evolution of supercritical
data beyond the formation of an event horizon. This allowed them to
determine
the ultimate behavior of the Yang-Mills hair which remains outside
the
black hole after the formation of an event horizon. 
In accordance with the
no-hair property it is expected that this hair is lost either via
radiation to
 infinity or by collapse to the black hole. Choptuik, Hirschmann, and
 Marsa
 found that the mechanism of loosing hair
 is different for black holes formed via two generalized (in the sense
explained in~\cite{chm}) types of collapse
 established in~\cite{ccb}. For a generalized supercritical Type II 
collapse most
of the hair is radiated off to infinity leaving the original black hole
 virtually unchanged. In contrast, for a generalized supercritical
 Type I
 collapse
 most of the hair
collapses and consequently the original black hole gets bigger.
 These two kinds of behavior are separated by the $n=1$
colored black hole (the static spherically symmetric black hole
solution of Einstein-Yang-Mills equations~\cite{cbh,vg}) which plays
the role
 of an
intermediate 
attractor
for near-critical solutions.  Similarly to the Type I collapse, the
lifetime $T$
of a near-critical solution staying in the vicinity of the intermediate
attractor was found to satisfy the scaling law
\begin{equation}
T \sim -\lambda \ln|p-p^{*}|\, ,
\end{equation}
where the coefficient $\lambda$ is the characteristic time scale for
the decay of the $n=1$ colored black hole, and $|p-p^{*}|$ is the
distance from the threshold ($p^{*}$) along a one-parameter family of
interpolating initial data.

 As described in~\cite{ccb},
the coefficient $\lambda$ can be
computed in two independent ways: either
directly from the nonlinear evolution, or perturbatively via linear
stability analysis. Since Choptuik, Hirschmann, and Marsa calculated
$\lambda$ only in the first manner, here we would like to compare
their
result with the predictions of linear stability analysis. 
We think that this comparison not only 
validates the results of Choptuik, Hirschmann, and Marsa but, in addition,
provides a useful test of the precision of
their numerical estimates.

We recall that according to the picture of critical solution as a
codimension-one attractor, the coefficient  $\lambda$ is a reciprocal
of the eigenvalue $\sigma$ corresponding to the single unstable mode,
{\em i. e.}, $\lambda=1/\sigma$.
 The single mode instability (within the magnetic ansatz) of
the n=1 colored black holes was established almost a decade
ago~\cite{sz,bw}, however there
are no (as far as we know) published results on how the eigenvalue 
of the unstable mode depends on the radius of the horizon (in
~\cite{ja} one of us determined this relationship in reference with
the issue of regularity
of the unstable mode at the horizon but no numerical values were
given there). We fill in this gap in the literature with Fig.~1,
where
the eigenvalue $\sigma$ is plotted as a function of the radius of the
horizon $r_h$
of $n=1$ colored black hole. This plot was generated by numerically
solving,
after Straumann and Zhou, their eigenvalue equation for small
perturbations~\cite{sz}.
\begin{figure}
\epsfxsize=9cm
\centerline{\epsffile{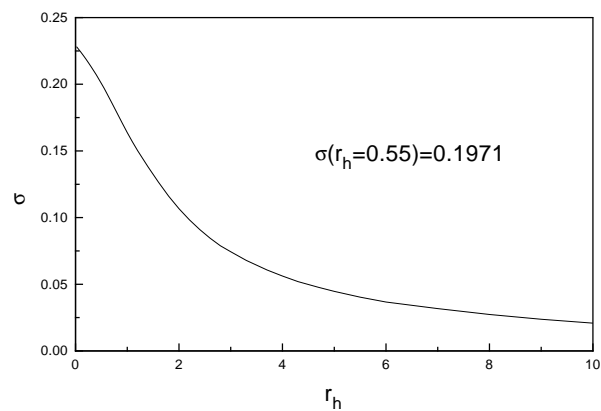}}
\caption{The eigenvalue $\sigma$ corresponding to the single unstable
  mode
about the $n=1$ colored black hole is
plotted against the radius of the event horizon $r_h$. As $r_h
\rightarrow 0$, $\sigma \rightarrow 0.2292$, the eigenvalue of the
unstable mode about the Bartnik-Mckinnon solution. In order to fix the
scale, the time coordinate is normalized to the proper time at infinity.}
\label{FIG1}
\end{figure}
Choptuik, Hirschmann, and Marsa estimated that $\lambda \approx 4.88$
 for
 $r_h \approx 0.55$ which compares well
with our calculation of $1/\sigma(0.55) =5.0736$. Unfortunately,
 this is the only value of $r_h$ for which we can make the comparison because
Choptuik, Hirschmann, and Marsa did not
estimate $\lambda$ for other values of $r_h$.
  It is worth noting that since the
eigenvalue $\sigma$ (and {\em eo ipso} the coefficient $\lambda$)
 varies with $r_h$, it is universal only among those 
interpolating families of initial
 data which share the {\em same} critical solution ({\em i. e.}, have
 the same $r_h$).

 Notice that, as follows from Eq.(1) and Fig.~1, 
for a given distance
 from the threshold, the lifetime $T$ increases with $r_h$, so 
colored black
 holes with large $r_h$ should be more clearly pronounced as
 intermediate attractors. However, this effect is partly offset by the
 fact that, in the limit of large $r_h$, colored black holes become
 almost
indistiguishable from the Schwarzschild black hole because the energy
of the Yang-Mills hair rapidly decreases with $r_h$.
In other words, the mass gap across the threshold is expected to
decrease with $r_h$. The upper bound for the mass gap is given
by
\begin{equation}
\Delta M (r_h) = M_{cbh}(r_h)-\frac{r_h}{2},
\end{equation}
which is the difference between the mass $M_{cbh}(r_h)$ of the $n=1$
colored 
black hole and the mass of the  Schwarzchild black hole with the same
radius of event horizon $r_h$. This maximal
mass gap would be achieved if all hair    remaining outside the horizon of
near-critical solutions would either {\em completely} disperse, or {\em
  completely} collapse. Since in the real collapse there is always a
fraction of hair which does otherwise, the actual mass gap is smaller
than $\Delta M$.
The maximal mass gap given
by Eq.(2) is plotted in Fig.~2.
\begin{figure}
\epsfxsize=9cm
\centerline{\epsffile{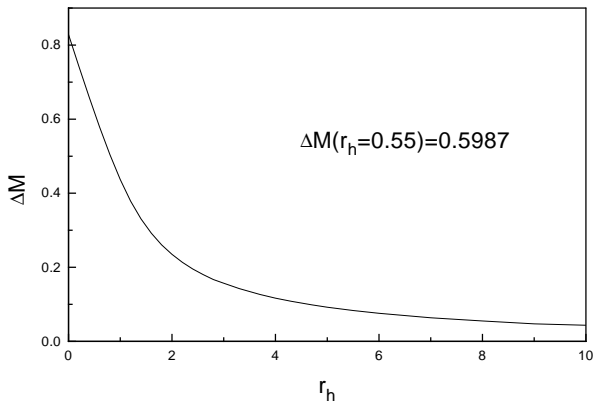}}
\caption{The maximal mass gap across the threshold as a function of
  the radius of the horizon. For $r_h=0.55$, Choptuik, Hirschmann, and
  Marsa obtain the mass gap $\sim
0.57$ which is approximately $95\%$ of the maximal mass gap.}
\label{FIG2}
\end{figure}
It would be interesting to compare Fig.~2 with the actual mass gap.
In particular, one might wonder whether the mass gap goes to zero for a
large but finite
$r_h$. If this really happened, then the line of colored black holes
on the phase diagram shown in Fig.~4 in~\cite{chm},  beginning at
the triple point for $r_h=0$, would terminate at a finite distance, in an
amusing similarity to the gas-liquid boundary on phase diagrams for
 typical substances.
We realize that the numerical verification of this speculation would be
rather difficult.

{\em Acknowledgments.} 
This research was supported  by the KBN grant
2 P03B 010 16.

\nopagebreak

\end{document}